\documentclass[journal]{IEEEtran}
\ifCLASSINFOpdf
\else
\fi

\usepackage{tikz}
\usepackage{amsmath}

\usepackage{filecontents}
\usepackage{amsmath}
\usepackage{algpseudocode}
\usepackage{algorithm}
\usepackage{hhline}
\usepackage{graphicx}
\graphicspath{ {figures/} }
\usepackage{svg}

\usepackage{booktabs} 
\usepackage{amsmath,amssymb}
\usepackage{todonotes}
\usepackage{url}
\usepackage{amsmath}
\usepackage{algpseudocode}
\usepackage{algorithm}
\usepackage{hhline}
\usepackage{graphicx}
\graphicspath{ {figures/} }
\usepackage{svg}
\usepackage{subfig}
\usepackage{float}
\usepackage{adjustbox}
\usepackage{tabularx}
\usepackage{balance}
\usepackage{url}
\usepackage{listings}

\begin{document}
%
\title{\textbf{GAIROSCOPE:} Injecting Data from Air-Gapped Computers to Nearby Gyroscopes}



\author{\IEEEauthorblockN{Mordechai Guri}\\
\IEEEauthorblockA{Ben-Gurion University of the Negev, Israel\\
	 Department of Software and Information Systems Engineering\\ Cyber-Security Research Center \\
Email: gurim@post.bgu.ac.il \\ Demo video: http://www.covertchannels.com}}


\IEEEoverridecommandlockouts
\makeatletter\def\@IEEEpubidpullup{6.5\baselineskip}\makeatother
\IEEEpubid{\parbox{\columnwidth}{
    A slightly modified version was accepted to 2021 18th International Conference on Privacy, Security and Trust (PST) ~\copyright~2021 IEEE
    DOI: 10.1109/PST52912.2021.9647842
     \\
}
\hspace{\columnsep}\makebox[\columnwidth]{}}

\maketitle

\begin{abstract}
It is known that malware can leak data from isolated, air-gapped computers to nearby smartphones using ultrasonic waves. However, this covert channel requires access to the smartphone's microphone, which is highly protected in Android OS and iOS, and might be non-accessible, disabled, or blocked. 

In this paper we present `GAIROSCOPE,' an ultrasonic covert channel that doesn't require a microphone on the receiving side. Our malware generates ultrasonic tones in the resonance frequencies of the MEMS gyroscope. These inaudible frequencies produce tiny mechanical oscillations within the smartphone's gyroscope, which can be demodulated into binary information. Notably, the gyroscope in smartphones is considered to be a 'safe' sensor that can be used legitimately from mobile apps and javascript. We introduce the adversarial attack model and present related work. We provide the relevant technical background and show the design and implementation of GAIROSCOPE. We present the evaluation results and discuss a set of countermeasures to this threat. Our experiments show that attackers can exfiltrate sensitive information from air-gapped computers to  smartphones located a few meters away via Speakers-to-Gyroscope covert channel. 
\end{abstract}

\section{Introduction}
Air-gap networks are isolated from the Internet to minimize the risk of data leakage. The air-gap isolation is enforced by removing wireless network interfaces (e.g., Wi-Fi), disabling USB devices, and controlling access to sensitive assets.  
Military networks such as the Joint Worldwide Intelligence Communications System \cite{Storefro90:online} are known to be isolated due to the classified information they hold. Other sectors such as critical infrastructure, governmental industries, and finance organizations are commonly storing their data within air-gapped networks \cite{Guri:2018:BAM:3200906.3177230}\cite{byres2013air}. In August 2021, Microsoft has announced the air-gapped Azure cloud region that is designed to handle top-secret data for the US intelligence community \cite{Microsof69:online}.

Nevertheless, despite a lack of direct connection to the outer world, air-gapped networks have been shown to be vulnerable to cyber-attacks. To breach isolated networks, attackers may compromise the supply chain, use malicious insiders, or employ social engineering attacks on employees. A publicized incident in which the air-gap was breached is the Stuxnet worm \cite{kushner2013real} which targets supervisory control and data acquisition (SCADA) systems and is believed to be responsible for causing damage to the nuclear facilities in Iran. ProjectSauron APT (Advanced Persistent Threat) discovered in 2016 breached air-gaps via an infected USB device that was used to leak data off the isolated network. Notably, this malware remained undetected for five years and used hidden partitions in the file system to hide its data. Several attacks on air-gapped facilities such as the power utilities \cite{Nobigdea65:online} and nuclear power plants \cite{AnIndian12:online} have been publicized in recent years.

\subsection{Covert Communication Channels}
Once the attacker breached the air-gapped network, he must use non-standard communication methods to exfiltrate the data outward. These methods are known as air-gap covert channels, or out-of-band covert channels \cite{Guri:2018:BAM:3200906.3177230}. Electromagnetic covert channels use electromagnetic emanation from various computer components to leak information \cite{guri2014airhopper,kuhn1998soft,kuhn2002compromising,vuagnoux2009compromising,guri2015gsmem}. Optical channels encode data over various sources of light such as LEDs and screens \cite{loughry2002information,Guri2017}. Thermal signals can also be used to maintain slow communication between systems \cite{guri2015bitwhisper}. Acoustic waves are a fundamental medium for data transfer, using inaudible sound. Acoustic methods have been explored in a wide range of studies over the years \cite{hanspach2014covert,deshotels2014inaudible,madhavapeddy2005audio,Guri2018Mosquito}. The existing acoustic methods suggest transmitting data through the air-gap via ultrasonic waves generated by computer loudspeakers. The inaudible sound encodes binary data that can be picked up by a nearby smartphone via its microphone. Note that these covert channels rely on the availability of a microphone in the receiving side. 

\subsection{Microphone Access in Mobile OS}
The acoustic covert channels require loudspeakers for the transmission and a microphone for the reception. However, microphones in mobile operating systems such as Android and iOS are considered protected sensors since they may seriously violate user privacy (e.g., by eavesdropping). Because of its importance, the application that uses the microphone must have permission to access the recording device and must ask the user for approval at run time. Since the access to microphones is highly restricted and monitored in the mobile OS, its use in an attack is a challenging task for the attacker. It mainly requires evading security mechanisms or exploiting vulnerabilities in the OS or applications.   

\subsection{Our Contribution}
In this paper, we introduce a new type of covert channel. Our malware uses ultrasonic sound waves to transmit data but does not require access to a microphone in the receiving smartphone device.
We show that malware can use the smartphone's gyroscope to receive the data covertly. Our method is based on the vulnerability of MEMS gyroscopes to specific ultrasonic frequencies, known as resonance frequencies. When this inaudible sound is played near the gyroscope, it creates an internal disruption to the signal output. The errors in the output can be used to encode and decode information. Note that gyroscopes in smartphones are considered as low permission sensors and hence can be used by applications with minimal permissions. Sensitive data can be modulated over the resonance frequencies and then received by a smartphone placed on the table. We implement and evaluate the covert channel discuss relevant countermeasures.

\subsection{Gyroscopes in Smartphones}
The speakers-to-gyroscope covert channel has several advantages for the attacker.

\begin{itemize}
	\item {\textbf{Permissions.}} Unlike microphones, gyroscopes are perceived as `safe' sensors. They are used by many types of applications to ease the graphical interfaces, and users may approve their access without suspicion.
	
	\item {\textbf{Visual indication.}} In Android and iOS, there may be no visual indication, notification icons, or warning messages to the user that an application is using the gyroscope, like the indications in other sensitive sensors. 
	\item {\textbf{Browser access.}} The gyroscopes may be accessed from HTML via standard JavaScript code. This implies that the attacker is not required to compromise the user's device or install a malicious application. Instead, the attacker can implant malicious JavaScript on a legitimate website that samples the gyroscope, receives the covert signals, and exfiltrates the information via the Internet.   
\end{itemize}

The rest of this paper is organized as follows: The attack model is discussed in Section \ref{sec:attack}. Related work is presented in Section \ref{sec:related}. Technical background is provided in Section \ref{sec:tech}. Sections \ref{sec:trans} and \ref{sec:rec}, respectively, contain details on data transmission and reception. Section \ref{sec:eval} present the evaluation. Defense and countermeasures are discussed in Section \ref{sec:counter}. We conclude in Section \ref{sec:conclusion}.

\section{Attack Model}
\label{sec:attack}
The attack model consists of transmitting computer(s) and receiving smartphone(s). The transmitter is executed in the air-gapped network, and the receiver is run on a  mobile phone or tablet belonging to workers or visitors.

\subsection{Air-gap Infection}
In the first stage of the kill chain, the attacker must breach the air-gapped network. Attackers may achieve it by attacking the supply chain or by using malicious insiders or deceived workers. Despite the high degree of separation from other environments, secure networks can be breached. The most famous cases are Stuxnet \cite{kushner2013real} and Agent.BTZ \cite{AgentBTZ65:online}, although other incidents are reported from time to time. In the case of Agent.BTZ in 2008, networks of the U.S. military base in the Middle East were infected via USB flash drive. The malware was delivered when this USB thumb drive was inserted into the laptop attached to United States Central Command. In 2018 the U.S. Department of Homeland Security (DHS) blamed Russian hackers for penetrating networks electric utilities in the state \cite{Nobigdea65:online}. Other malware and attack tools like and Ramsay \cite{ESETRese49:online} and USBFerry \cite{Hackerst65:online}, that target air-gapped networks, were found in the wild.

\subsection{Mobile Infection}
In addition, the smartphones of employees are infected with a rough application. Mobile devices can be infected through phishing, by using social engineering techniques, via malicious email attachments, compromised websites, Wi-Fi, malicious advertisements, and other attack vectors \cite{provos2007ghost, cova2010detection, sood2011malvertising, peltier2006social, smutz2012malicious}. 


%
%
%

\subsection{Collection}
Having a foothold in the organization, malware in the compromised network gathers information of interest, such as encryption keys, key-logging information, credentials, access codes, etc.

\subsection{Exfiltration} 
In the exfiltration phase, the malware encodes the data and broadcast it to the environment, using covert acoustic sound waves in the resonance frequency generated from the computer's loudspeakers. A nearby infected smartphone `listens' through the gyroscope, detects the transmission, demodulates and decodes the data, and transfers it to the attacker via the Internet (e.g., over Wi-Fi).

\begin{figure}
	\centering
	\includegraphics[width=\linewidth]{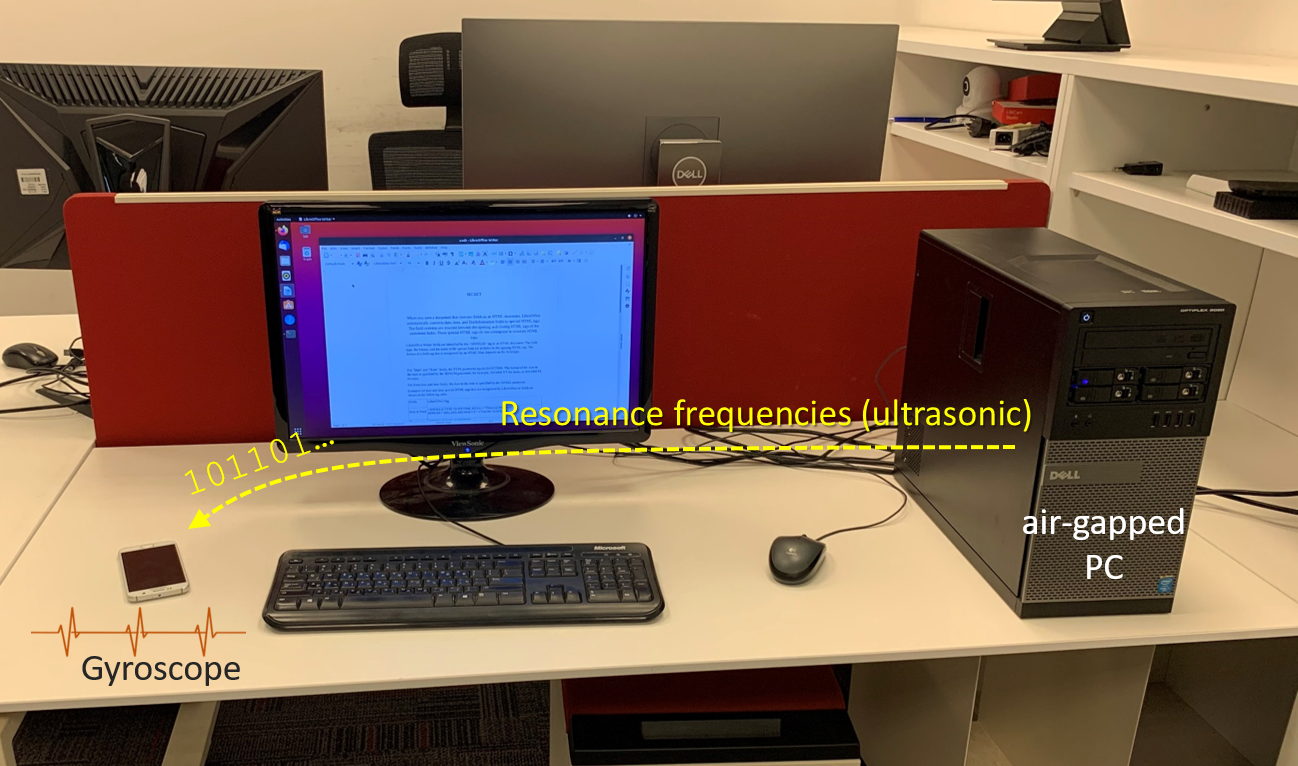}
	\caption{\textbf{Attack scenario.} The air-gapped workstation broadcasts binary data modulated on top of ultrasonic waves in the resonance frequencies that oscillate the nearby MEMS gyroscope. The application in the smartphone samples the gyroscope, demodulates the signal, and transmits the decoded data to the attacker through Wi-Fi. Note that microphone access is not required.} 
	
%
	\label{fig:scenario}
\end{figure}

The chain of attack is illustrated in Figure \ref{fig:scenario}. The air-gapped workstation broadcasts data modulated on top of ultrasonic waves in the resonance frequencies that oscillates the nearby MEMS gyroscope. The application in the smartphone samples the gyroscope, demodulates the signal, and transmits the decoded data to the attacker through Wi-Fi.

\section{Related Work}
\label{sec:related}
Covert channels, in general, have been widely covered in the professional literature for years \cite{giani2006data,murdoch2005embedding,zander2007survey}.
Air-gap covert channels are used when no network connectivity is available for the attacker. These communication methods can be mainly classified into electromagnetic, magnetic, optical, thermal, and acoustic. Many prior studies proposed the use of electromagnetic radiation for data leakage. The radio frequencies (RF) emanated from electronic components in the computer \cite{kuhn1998soft, guri2017bridging,funtenna86:online,Guri2016}. Guri et al. exploit video cards in air-gapped computers to transmit covert FM signals \cite{guri2017bridging}. Later on, GSMem and AIR-FI malware could generate electromagnetic signals from air-gapped systems in the cellular and Wi-Fi frequency bands, respectively. These frequencies are used to modulate and exfiltrate data. USBee \cite{guri2016usbee} introduced attack scenarios in which attackers use different sources of electromagnetic radiation from an air-gapped computer's motherboard for data exfiltration. Magnetic fields emitted from a computer CPU can be used to leak data from Faraday caged air-gapped computers \cite{guri2019odini,guri2018magneto}. PowerHammer is another method in which malware on air-gapped computers exfiltrates data through electromagnetic emissions encoded on the power lines \cite{guri2019powerhammer}. 
Several studies have proposed the use of optical emanation for air-gap communication. Loughry and Guri demonstrated how to exploit the status indicators in the keyboard LEDs for data exfiltration. \cite{loughry2002information}\cite{guri2019ctrl}. The hard disk drive (HDD) LEDs \cite{Guri2017}, router and switch LEDs \cite{guri2018xled}, and security cameras and their IR LEDs \cite{guri2019air}, also proposed as methods for exfiltrating data from air-gapped networks. Researchers also showed how to exfiltrate data from air-gapped computers via screen brightness \cite{guri2019brightness}, and hidden images projected on the screen \cite{guri2016optical}. BitWhisper is a unique covert channel based on the thermal medium \cite{guri2015bitwhisper}. It enables two-way communication between air-gapped computers by encoding data on temperature changes.

\subsection{Acoustic}
In acoustic covert channels, data is transmitted via sound waves in various frequencies. In most of the previous work, the sound waves are generated from the computer loudspeakers. A Survey of audio communication was given in \cite{madhavapeddy2005audio}. Carrara and Adams discussed the threat of acoustic covert channel between air-gapped systems \cite{carrara2014acoustic}. Hanspach used high-frequency sound waves to maintain communication between laptops. In 2018, researchers presented MOSQUITO \cite{Guri2018Mosquito}, a covert communication channel between two air-gapped computers without microphones via speakers-only communication. The concept of communicating over inaudible sounds has been extended for different scenarios using smartphone devices \cite{deshotels2014inaudible}. The acoustic methods presented above require speakers to produce noise. Researchers also introduced a few methods which do not require loudspeakers. Fansmitter, Diskfiltration, and CD-LEAK are three malware that maintains the exfiltration of data from an air-gapped computer via noise intentionally emitted from the PC fans, HDD, and CD/DVD, respectively \cite{guri2020fansmitter}\cite{guri2017acoustic} \cite{guri2020cd}. Recently, researchers showed that the power supply of computers and devices could be used to generate sonic and ultrasonic waves for data exfiltration \cite{guri2020power}.
Yonezawa \cite{yonezawa2011transferring} and Deshotels \cite{Deshotels2014} showed that mobile devices with vibrator components could produce vibrations that generate low volume sound. They showed that the vibrations could be sampled and decoded via an accelerometer in mobile phones. Note that these covert channels are not relevant to real air-gap environments that have no vibrators or accelerometers. Vibrational covert channels are also discussed by Hasan et al. \cite{hasan2013sensing} and Matyunin et al. \cite{matyunin2019vibrational}. They showed that the mobile phone could receive low-frequencies audio generated with a sub-woofer attached to the transmitting computer. However, the requirement for sub-woofers in the compromised computer limits the attack model since sub-woofers are not commonly used in workstations. Our method uses ordinary loudspeakers to transmit ultrasonic signals and doesn't require a smartphone microphone on the receiving side. Instead, it uses the low permission gyroscope sensor to receive the information. Table \ref{tab:comp} compares the acoustic communication channels.

\begin{table*}[]
		\centering
        \caption{Comparison with the existing acoustic covert channels}
        \label{tab:comp}
	\begin{tabular}{@{}lllllllll@{}}
		\toprule
		\# & Method         & Transmitter     & Receiver        & Speakers    & Frequencies     & Ultrasonic & Max speed            & Max distance \\ \midrule
		1  & MOSQUITO       & PC speakers     & Microphones & Required & 0-24Khz         & Yes        & \textless 1k bit/sec & 8 m          \\
		2  & Ultrasonic     & PC speakers     & Microphones & Required  & 19-24 KHz       & Yes        & \textless 1k bit/sec & 11 m         \\
		3  & Low frequency  & PC speakers     & Microphones & Required  & 0-50 Hz         & No         & 2 bit/sec            & 3 m          \\
		4  & Fansmitter     & PC fans         & Microphones & Not required            & 0-300 Hz        & No         & 1 bit/sec            & 7-8 m        \\
		5  & Diskfiltration & Hard Disk Drive & Microphones & Not required            & 2000 Hz         & No         & 2 bit/sec            & 2 m          \\
		6  & CD-LEAK        & CD/DVD drive    & Microphones & Not required            & \textless 5 KHz & No         & 1 bit/sec            & 1.5 m        \\
		7  & Power-Supplay  & Power supply    & Microphone & Not required            & 0 - 24 KHz      & Yes        & 60 bit/sec              & 6 m  
		\\
		9  & AiR-ViBeR  & PC fans    & Accelerometer & Not required            & 0-500 Hz      & Yes        & 1 bit/sec              & 2 m          
		\\ 
		9  & GAIROSCOPE  & PC speakers   & Gyroscope & Required           & above 18 kHz      & Yes        & 8 bit/sec              & 8 m
		\\ 
		\bottomrule
	\end{tabular}
\end{table*}

\section{Technical Background}
\label{sec:tech}
In this section, we provide the technical background on the gyroscope sensors in smartphone devices. A Gyroscope sensor is used to determine the reference direction of the device in three dimensions (X, Y, and Z). It provides stability in mobile navigation, stabilizes the user interface, rotates the screen on handled devices, etc. A Gyroscope sensor is present in all modern smartphones to sense angular rotational velocity and acceleration. It helps to augment human motion interactive games and ease the user interface applications where the screen content is rendered due to the presence of a gyroscope (`autorotation').

There are various gyroscopes sensors based on different types of technology, such as MEMS gyroscopes, solid-state ring lasers, fiber optic gyroscopes, and sensitive quantum gyroscopes. The gyroscope sensors integrated into smartphones belong to the class of  Micro-electromechanical systems (MEMS) gyroscopes. The MEMS gyroscope output the ate of turn (rad/s) without a fixed point of reference. Internally, MEMS gyroscopes are constructed of a vibrating object with multiple-axis freedom.

The gyroscope sensor in mobile devices (e.g., Android and iOS) is accessed through the sensors API. The output vector is given in radians/second, and it measures the rate of rotation around the device's local X, Y, and Z axes. The rotation is positive in the counter-clockwise direction. E.g., in Android OS, the returned values[0], values[1], and values[2] contain the angular speed around the X-axis, Y-axis, and Z-axis, respectively. The applications usually integrate the output values over time to calculate a rotation over the time step.

It is known that acoustic tones degrade MEMS sensors in a frequency range knowns as the `resonance frequencies. This ultrasonic input produces erroneously low-frequency angular velocity readings in the X, Y, or Z direction(s) \cite{khazaaleh2019vulnerability}. The vulnerability of MEMS sensors to ultrasonic corruption is due to the mechanical structure of a MEMS gyroscope. For example, as observed by \cite{khazaaleh2019vulnerability} the misalignment between the driving and sensing axes is one of the main causes of the false output generated by the gyroscope. The phenomenon and its physical and mechanical roots are discussed in relevant literature \cite{algamili2021review}. It was observed in the previous works that the typical resonance frequencies of MEMS are within a fragmented band in the ultrasonic frequencies range mainly above 18 kHz. The frequency of the resulting vibrations within the sensor is determined by the structure of the MEMS gyroscope, it's positioning, and the distance from the sound source \cite{algamili2021review}\cite{son2015rocking}\cite{khazaaleh2019vulnerability}. 

Studies in recent years focused on using the vulnerability of MEMS sensors to ultrasonic waves for attack purposes. Khazaaleh et al. discuss the general vulnerability of MEMS gyroscopes to
targeted acoustic attacks \cite{khazaaleh2019vulnerability}. Son et al. introduced attacks on drones navigation systems via sound noise attacks on gyroscopic sensors \cite{son2015rocking}. Tu et al. discussed the malicious control over
situation systems by spoofing inertial sensors such as MEMS gyroscopes \cite{tu2018injected}.


\section{Transmission}
\label{sec:trans}
This section describes the signal generation, data modulation, and transmission protocol.

\subsection{Signal Generation}
Sound waves in the range of resonance frequency generate disruptive vibrations in the gyroscope's internal components. The disruptive signal is measured by the frequency of changes in the angular velocity output of the gyroscope sensor. The effect on the angular velocity can be measured in a single axis, two, or all three axes. Figure \ref{fig:XYZ} shows the effect of the resonance frequency on the output of the gyroscope sensor in the Samsung Galaxy S10 device. In this case, the transmitter generates sound in frequencies of 19067 Hz and 19065 Hz. As can be seen, the affected output is mainly in the X-axis and Y-axis.

To generate a signal, we play an acoustic wave in the specific resonance frequency for a certain amount of time. The following python script shows the code to build the sinus wave in 19000 Hz for 0.5 seconds with a volume of 20\%.

\begin{figure}[ht!]
	\centering
	\includegraphics[width=1.0\linewidth]{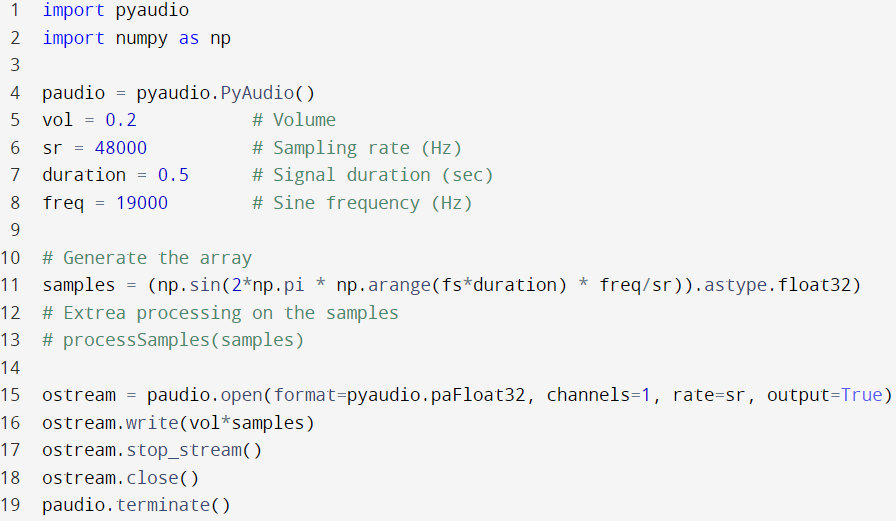}
	\label{fig:code}
\end{figure}

\begin{figure}
	\centering
	\includegraphics[width=1.0\linewidth]{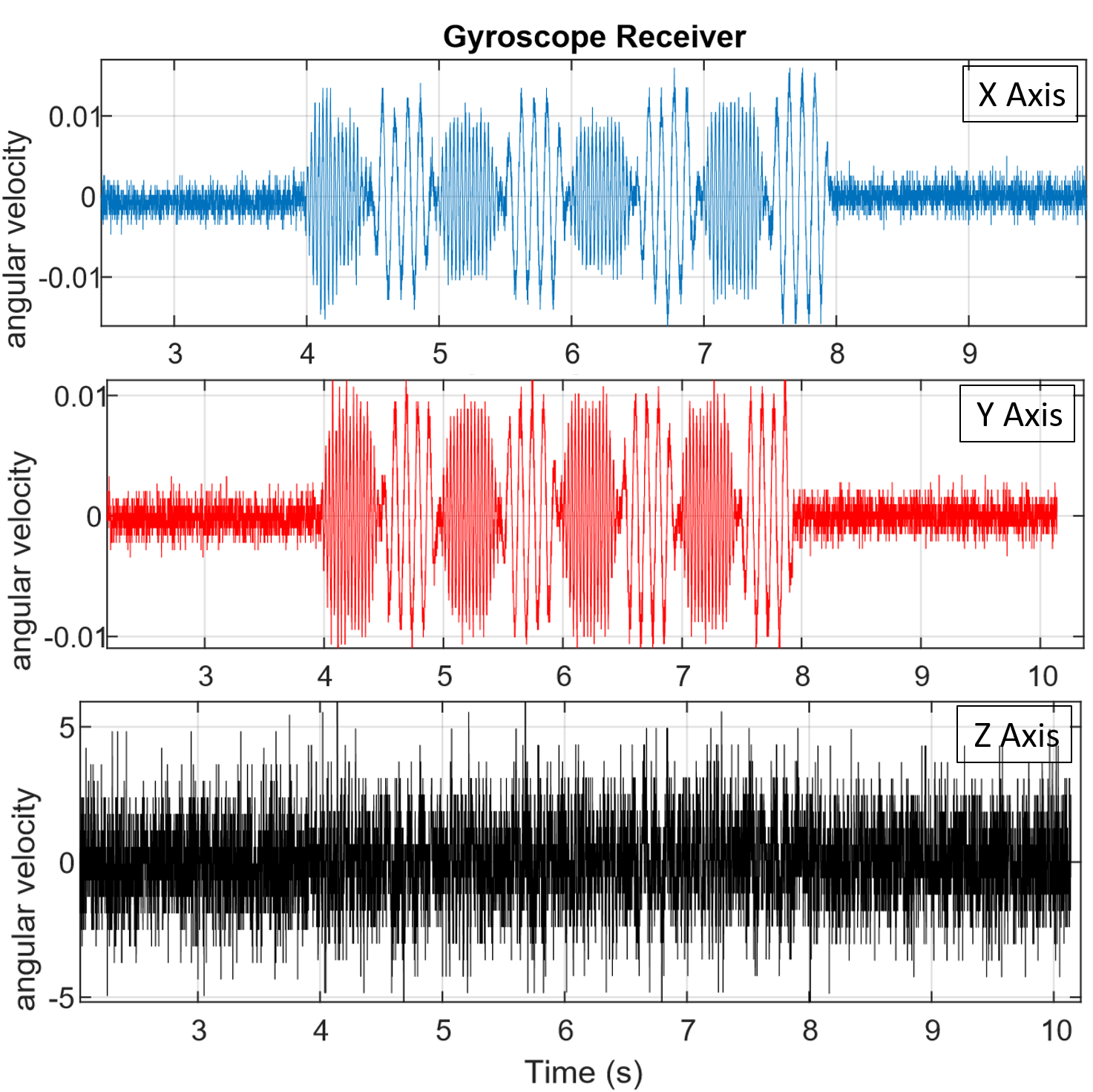}
	\caption{The effect of the resonance frequency on the output of the gyroscope sensor in the Samsung Galaxy S10 device. The affected output is mainly in the A-axis and Y-axis.}
	\label{fig:XYZ}
\end{figure}

\subsection{Frequency-Shift Keying}
In frequency-shift keying (FSK), the data is represented by a change in the frequency of a carrier wave. Recall that the transmitting code can determine the frequency of the signal generated in the gyroscope. In FSK, each frequency represents a different symbol.

\begin{figure}
	\centering
	\includegraphics[width=\linewidth]{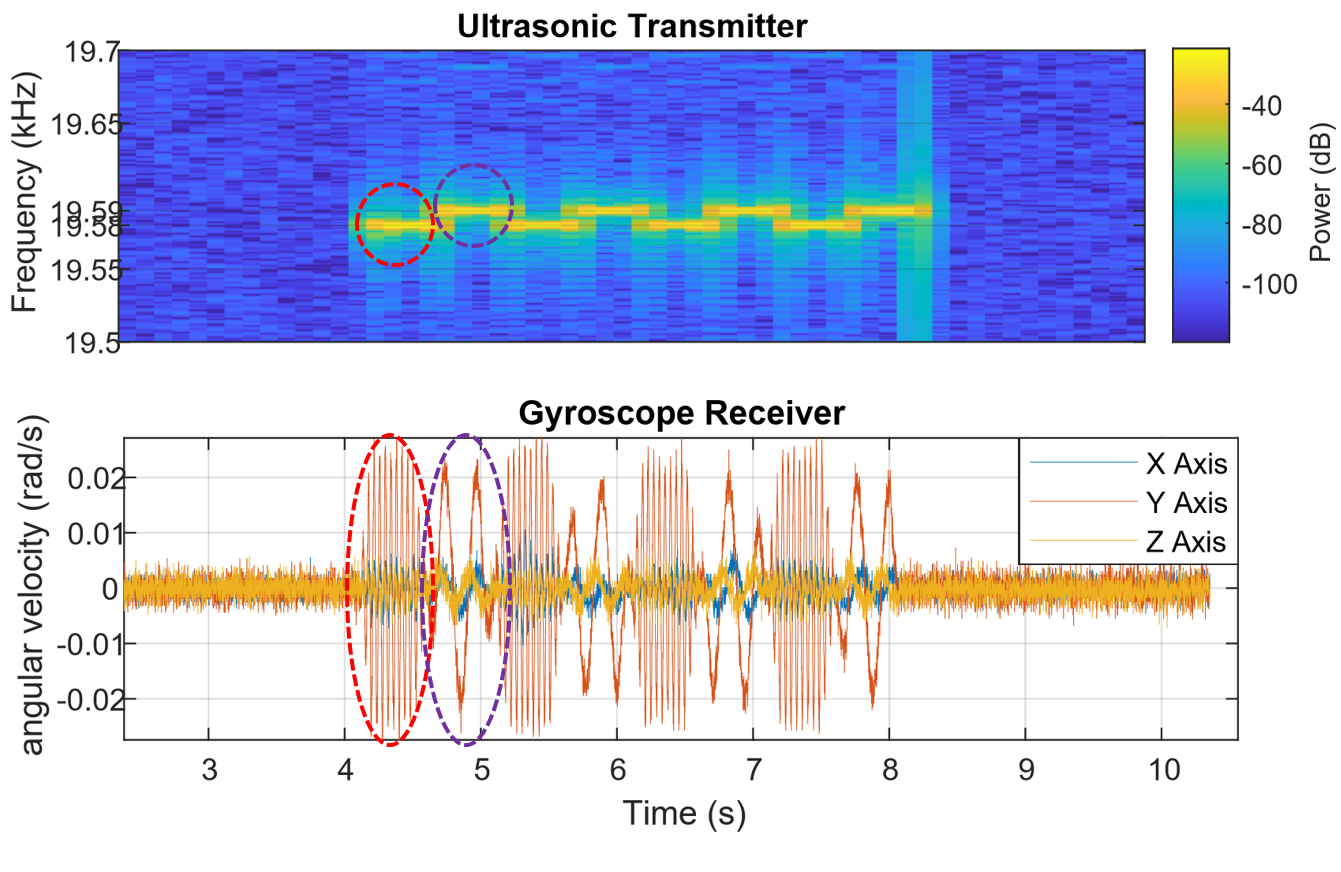}
	\caption{Spectrogram of the B-FSK modulation as generated by the computer (top) and received by the smartphine (bottom).}
	\label{fig:FSKSPEC}
\end{figure}

Figure \ref{fig:FSKSPEC} shows the time-frequency spectrogram of a binary sequence (`01010101') modulated with two resonance frequencies (B-FSK) as transmitted from a PC. The frequencies 10956 Hz and 19059 Hz in this modulation have been used to encode the symbols `0' and `1', respectively.

%
%

%

\subsection{Transmission Protocol}
We transmit the data in small frame packets composed of a preamble, a payload, and a parity bit for error detection.

\begin{itemize}
	\item \textbf{Synchronization.} A synchronization header is transmitted at the beginning of the frame. It consists of a sequence of four bits (`1010') which determine the beginning of a transmission and the properties such as amplitude and frequency. 
	\item \textbf{Payload.} The payload is the raw data of 12 bits to be transmitted. Note that the payload size of 12 bits was arbitrarily chosen. A more complex frame may contain a dynamic size of the payload. 
	\item \textbf{Parity bit.} For error detection, a parity bit is added to the end of the frame. The receiver calculates the parity for the whole received payload (excluding the synchronization bits), and if it differs from the received parity bit, an error is detected. 
\end{itemize}

%

\section{Reception}
\label{sec:rec}
We implemented a GAIROSCOPE receiver as an application for the Android OS based on the B-FSK modulation. The receiver app samples the gyroscope sensor and performs signal processing. The demodulator application is described in Algorithm \ref{alg:a3}.

The application continuously samples and processes the gyroscope sensor output. It handles data sampling, signal processing, and data demodulation. In our case, we sampled the gyroscope sensor using the Android sensors API. We used the type TYPE\_GYROSCOPE to sample the rate of rotation in three axes. It then performs a fast Furrier transform (FFT) to measure the FSK frequencies $f_0$ and $f_1$. The value of the headers bits determines the frequency and amplitude of the payload.

\begin{algorithm} 
	
	\caption{demodulate($bitTime$, $f_{0}$, $f_{1}$, $sampleRate$)} 
	
	\begin{algorithmic}[1] 
		\State $ onSensorChanged(SensorSamples\ samples)\  $
		\State $ sync \gets false $
		\State $ lockSamples \gets true $
		\State $ index f_{0} \gets fftSize * f_{0}/sampleRate $
		\State $ index f_{1} \gets fftSize * f_{1}/sampleRate $
		\State $ samplesPerBit \gets sampleRate*bitTime/(fftSize - noverlap) $
		
		\State $ magnitude \gets getVectorMagnitude(samples) $
		\State $ buffer.append(magnitude)$
		\If{$ buffer.size() == fftSize$}
		\State $ fftWindow \gets fft(buffer) $
		\State $ buffer.removeRange(0, fftSize - noverlap)$
		\\
		\State $ amplitude f_{0} \gets abs(fftWindow[index f_{0}]) $
		\State $ amplitude f_{1} \gets abs(fftWindow[index f_{1}]) $
		\\
		\If {$amplitude f_{0} > amplitude f_{1}$}
		\State $samples.append(0)$
		\Else
		\State $samples.append(1)$
		\EndIf
		\\  
		\If {$not sync$}   
		\State $sync = detectSynch(samples[X],samples[Y],samples[Z])$
		\If {$sync = true$}    
		\State $ setAffectedAxed(sync)$
		\State $ sync = true$
		\EndIf
		\EndIf
		\If {$enabled$}
		\State $state = "DEMODULATE" $
		\State $sync = demodulate(buffer,12,sync)$
		\State $valid = checkParity(buffer[parityIndex])$
		\State $state = "SYNC" $
		\EndIf	
		\EndIf
	\end{algorithmic}
	
	\label{alg:a3}
\end{algorithm}

\subsubsection{Signal Processing} The first step is to measure the signal and transfer it to the frequency domain. This step includes performing a fast Fourier transform (FFT) on the sampled signal, given the parameters $sampleRate$ and $fftSize$. The signal measured is stored in a buffer after applying a filter for noise
mitigation. This data is used later in the demodulation routines. The noise mitigation function is applied to the current sample by averaging it with the last $w$ samples. Note that in our case, the FFT is done for the samples of the three axes (X-axis, Y-axis, and Z-axis).

\subsubsection{Synch Detection} In the $SYNC$ state, the receiver
searches for a sequence to identify a frame header. If the synchronization sequence '1010' is detected, the
state is changed to $DEMODULATE$ to initiate the
demodulation process. Based on the preamble sequence, the receiver determines the channel parameters, signal time ($T$), and frequencies $f_{0}$ and $f_{1}$. Note that in this phase, we are scanning each of the axes to determine which is/are affected by the resonance frequencies.

\subsubsection{Demodulation} In the $DEMODULATE$ state, the payload is
demodulated given the signal parameters retrieved in the
$SYNC$ state. The demodulated bit is added to the current payload. When the payload of bits and the parity bit are received, the algorithm returns back to the preamble detection state ($SYNC$). 

\begin{figure}
	\centering
	\includegraphics[width=\linewidth]{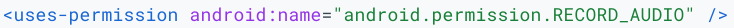}
	\caption{Spectrogram of a full frame modulated with B-FSK using the resonance frequencies, as received by a nearby smartphone.}
	\label{fig:frame}
\end{figure}

A spectrogram of a full-frame transmitted from a computer and received by a nearby smartphone is presented in Figure \ref{fig:frame}. In this case, B-FSK was used for modulation. The frame includes the synchronization bits (`1010'), the payload (`110100110100'), and the parity bit (`1').

\subsection{Sensor Permissions}
Mobile OS such as Android and iOS protect sensitive sensors such as microphones from illegal access via architectural features and security mechanisms. To record in Android OS, a recording permission tag must be included in the app's manifest file. However, the gyroscope is considered a legitimate and safe sensor, and mobile OS Android and iOS do not request high permissions to read the outputs of its samples. In addition, no visual indication such as a notification icon is shown during the application access to the gyroscope. Furthermore, the gyroscope may be accessed from a Web browser via standard JavaScript code. The attacker can implant malicious JavaScript on legitimate websites that sample the gyroscope's data using the \texttt{DeviceOrientationEvent} or other APIs. The following sample of code uses the Sensor APIs from a web page to read the angular velocity of the device along all three axes \cite{Sensoron5:online}. This API is supported by Chrome Android, WebView Android, and Opera Android.  

\begin{figure}[ht!]
	\centering
	\includegraphics[width=1.0\linewidth]{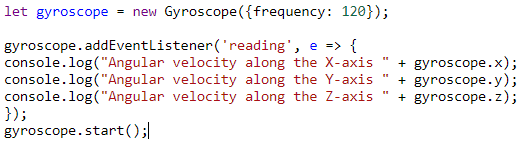}
	\label{fig:code2}
\end{figure}

\section{Evaluation}
\label{sec:eval}
This section provides an evaluation of the acoustic waveforms generated with the GAIROSCOPE transmitter. We also present the experimental setup and discuss the measurements and results.

\begin{table*}[]
	\label{tab:rec111}
	\centering
	\caption{Mobile receivers used for the evaluation}
	\begin{tabular}{@{}lllll@{}}
		\toprule
		\#                 & Device                           & Operating System & Gyroscope & Resonance Frequencies \\ \midrule
		OnePlus 7          & OnePlus 7 256GB 8GB RAM          & Android 9.0      & LSM6DSM   & 20,120 Hz, 20,135 Hz  \\
		Samsung Galaxy S9  & Samsung Galaxy S9 128GB 8GB RAM  & Android 10.0     & LSM6DSL   & 19,580 Hz, 19,588 Hz  \\
		Samsung Galaxy S10 & Samsung Galaxy S10 128GB 8GB RAM & Android 10.0     & LSM6DSM   & 19,065 Hz, 19,077 Hz  \\ \bottomrule
	\end{tabular}
\end{table*}

\subsection{Measurement Setup}
\label{sec:measurement_setup}

For the transmission, we used DELL Latitude 7490 laptop with Core i7 CPU, 16GB of RAM, and 512GB SSD with Linux Ubuntu 18.04 LTS 64-bit. For the reception, we used three types of mobile receivers: OnePlus 7, Samsung Galaxy S9, and Samsung Galaxy S10.


The three mobile receivers are specified in Table III. The devices have an integrated LSM6DSO chip, which is a system-in-package featuring a digital accelerometer and digital gyroscope. In the evaluation, we used the B-FSK modulation to transmit binary data. The two frequencies used to modulate `0' and `1' for each device are specified in the `Resonance Frequencies' column.

During the tests, the acoustic signals were recorded with three smartphones running the Android OS. The gyroscope sampling was performed using the smartphone's internal gyroscope at a sampling rate of 500 Hz. We used the \textit{MathWorks MATLAB} environment for signal processing and the \textit{Praat} framework for additional spectral analysis \cite{Praatdoi67:online}.

The experiments took place in a standard lab room (4.0 x 9.0 meters) with an active HVAC and during daily working hours. 

%
\begin{figure}[]
	\centering
	\includegraphics[width=1\linewidth]{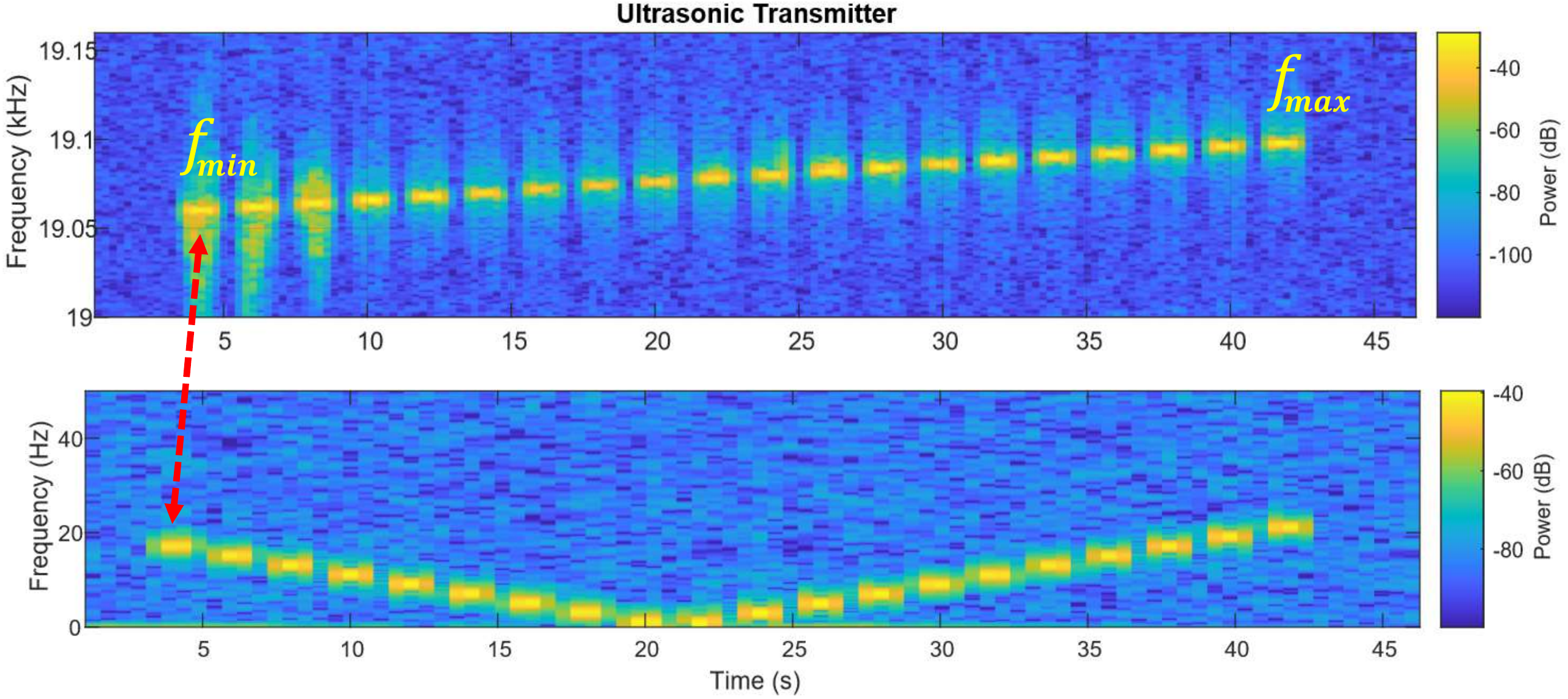}
	\label{fig:sweep1}
	\caption{A chirp signal on the resonance frequencies as transmitted acoustically (top spectrogram) and sampled by the gyroscope output (bottom spectrogram).} 
\end{figure}

\subsection{Resonance Frequencies}
In order to find the resonance frequencies bands for each device, we run a battery of measurements. We generated slow `up-chirp' signals (sweeps) from a transmitter computer while sampling the gyroscope output in the X, Y, and Z-axis for each of the mobile devices. We then mapped the corresponding resonance frequencies within the gyroscope sensor. 
Figure 5 depicts the transmitter sweep signal and the corresponding affected output of the gyroscope for the Samsung Galaxy S9 receiver. The range of resonance frequencies and the corresponding angular velocities are given in Table IV.

\begin{table}[]
	\centering
	\label{tab:ranges1}
	\caption{Resonance frequency measurements}
	\begin{tabular}{@{}lll@{}}
		\toprule
		Device             & Resonance frequencies & Angular Velocity Freq. \\ \midrule
		OnePlus 7          & 20050-2200 Hz   & 0-100 Hz                        \\
		Samsung Galaxy S9  & 19500 Hz - 19660 Hz   & 0-100 Hz Hz                        \\
		Samsung Galaxy S10 & 19000 Hz - 19077 Hz   & 0-80 Hz Hz                        \\ \bottomrule
	\end{tabular}
\end{table}

\subsection{Measurements Room \#1}
We measured the signal-to-noise (SNR) ratio and the bit error rates (BER) of the communication in an open space lab room. We transmitted a sequence of bits modulated with B-FSK and used the receiver app in the mobile device to sample, record, measure, and demodulate the signal. In the experiment, we used 700 ms per bit (1.4 bit/sec) and 125 ms per bit (8 bit/sec) for shorter distances of up to 100 cm.
The results of the SNR and BER are given in Tables \ref{tab:SNR1} and \ref{tab:BER1}, respectively. As can be seen, the Samsung Galaxy S9 and S10 could demodulate the transmitted data to a maximum distance of 150 cm and 200 cm, receptively. The OnePlus receiver reached a limited distance of 100 cm. Faster transmission rates yield a weak and unstable signal with low SNR and high BER that are impractical for reliable communication. This time limit is mainly due to the mechanical nature of the resonance frequencies and the disruption effect on the receiving gyroscope.

\begin{table}[]
	\centering
	\caption{SNR in room \#1}
	\label{tab:SNR1}
	\begin{tabular}{@{}llllll@{}}
		\toprule
		Device             & 0 cm & 50 cm & 100 cm & 150 cm             & 200 cm             \\ \midrule
		OnePlus 7          & 35 dB  & 25 dB   & 15 dB    & 10 dB & 7 dB \\
		Samsung Galaxy S9  & 43 dB  & 33 dB   & 33 dB    & 30 dB               & 25 dB                \\
		Samsung Galaxy S10 & 37 dB  & 33 dB   & 28 dB    & 20 dB            & 10 dB \\ \bottomrule
	\end{tabular}
\end{table}

\begin{table}[]
	\centering
	\caption{BER in room \#1}
	\label{tab:BER1}
	\begin{tabular}{@{}llllll@{}}
		\toprule
		Device             & 0 cm & 50 cm & 100 cm & 150 cm             & 200 cm             \\ \midrule
		OnePlus 7          & 0\%  & 0\%   & 0\%    & \textgreater{}20\% & \textgreater{}20\% \\
		Samsung Galaxy S9  & 0\%  & 0\%   & 0\%    & 0\%                & 0\%                \\
		Samsung Galaxy S10 & 0\%  & 0\%   & 0\%    & 12.5\%             & \textgreater{}20\% \\ \bottomrule
	\end{tabular}
\end{table}

\subsection{Measurements Room \#2}
As can be observed in the measurements above, in open space rooms, the effective range of resonance frequencies transmitted from the computer was limited. This is due to the propagation of the acoustic waves. We measured the signal-to-noise (SNR) ratio and the bit error rates (BER) of the communication in a narrow room (2.0 x 9.0 meters), where the wave propagation is minimized. We transmitted a sequence of bits modulated with B-FSK and used the receiver app in the mobile device to sample, record, measure, and demodulate the signal. In the experiment, we used 700 ms for each bit (1.4 bit/sec). The results of the SNR and BER are given in Tables \ref{tab:SNR1} and \ref{tab:BER1}, respectively. As can be seen, the Samsung Galaxy S9 and S10 could demodulate the transmitted data to a maximum distance of 800 cm. Note that Galaxy S10 reception was fragmented and did not work well at some distances. The OnePlus receiver reached a limited distance of 200 cm.

\begin{table*}[]
	\centering
	\caption{SNR in room \#2}
	\label{tab:SNR2}
	\begin{tabular}{@{}lllllllll@{}}
		\toprule
		Device             & 100 cm & 200 cm & 300 cm & 400 cm & 500 cm & 600 cm & 700 cm & 800 cm \\ \midrule
		OnePlus 7          & 26 dB  & 20 dB  & 15 dB  & 14 dB  & 13 dB  & 10 dB  & 14 dB  & 12 dB  \\
		Samsung Galaxy S9  & 22 dB  & 27 dB  & 20 dB  & 16 dB  & 28 dB  & 16 dB  & 18 dB  & 26 dB  \\
		Samsung Galaxy S10 & 18 dB  & 15 dB  & 15 dB  & 14 dB  & 18 dB  & 10 dB  & 10 dB  & 18 dB  \\ \bottomrule
	\end{tabular}
\end{table*}

\begin{table*}[]
	\centering
	\caption{BER in room \#2}
	\label{tab:BER2}
	\begin{tabular}{@{}lllllllll@{}}
		\toprule
		Device             & 100 cm & 200 cm & 300 cm             & 400 cm             & 500 cm             & 600 cm             & 700 cm             & 800 cm             \\ \midrule
		OnePlus 7          & 0\%    & 0\%    & \textgreater{}20\% & \textgreater{}20\% & \textgreater{}20\% & \textgreater{}20\% & \textgreater{}20\% & \textgreater{}20\% \\
		Samsung Galaxy S9  & 0\%    & 0\%    & 0\%                & 0\%                & 0\%                & 0\%                & 0\%                & 0\%                \\
		Samsung Galaxy S10 & 0\%    & 0\%    & \textgreater{}20\% & 12.5\%             & 0\%                & \textgreater{}20\% & \textgreater{}20\% & 0\%                \\ \bottomrule
	\end{tabular}
\end{table*}


\subsection{Measurements with multiple frequencies}
We also used the FSK modulation with multiple frequencies to test the range of the covert channel when more than two frequencies are used. In this case, we used a maximum of 32 frequencies to encode different symbols (5 bits). In the experiment, we used the range of 19000 Hz -19105 Hz to generate different resonance frequencies within the receiver. The results of BER are given in Tables \ref{tab:BER3}. As can be seen, the Samsung Galaxy S10 and the OnePlus 7 could demodulate the transmitted data to a maximum distance of 600 cm.
\begin{table}[]
	\caption{BER with multiple frequencies}
	\label{tab:BER3}
	\begin{tabular}{@{}lllllll@{}}
		\toprule
		Device             & 1 m & 2 m & 3 m & 4 m & 5 m & 6 m \\ \midrule
		OnePlus 7          & 0\% & 0\% & 0\% & 0\% & 0\% & 0\% \\
		Samsung Galaxy S10 & 0\% & 0\% & 0\% & 0\% & 0\% & 0\% \\ \bottomrule
	\end{tabular}
\end{table}

\subsection{Threat Radius}
The above results indicate that the covert channel can be used to transfer data with bit rates of 1-8 bit/sec at distances of 0 - 600 cm, with B-FSK or M-FSK modulations. In certain rooms (e.g., narrow rooms), the transmitter can reach 800 cm. The method could be used to transfer a small amount of data, such as short texts, encryption keys, passwords, or keystroke logging. Note that previous work \cite{guri2017bridging} indicates that employees used to place their mobile phones close to their workstations on the desk.

\subsection{Vibrational Interference}
We check whether the mechanical vibrations generated by the PSU, chassis, and CPU fans from desktop computers can interfere with the signal and affect the output of the gyroscope. 
The power supply fans are integrated into the power supply at the back of the case, and their speed is static. The chassis fans are installed at the back of the computer case, and their speed is automatically changed due to the internal heat. The CPU fans are mounted on top of the CPU socket. It cools the CPU's heatsink. The speed is regulated by a controller that adjusts the fan speed according to the current CPU workload. They are working in typical ranges of 900-4500 rotations per minute. Our tests show that the covert channel signals are not affected by the vibrations generated by the PSU, chassis, and CPU fans.

\subsection{Acoustic Interference}	
Several external sources of acoustic noise can interfere with the resonance frequencies used in this covert channel and override them. We conducted experiments and mapped the sources of potential interference from various objects in our lab. The list is given in Table \ref{tab:inter}. Notably, most of the measured environmental noise was within the audible band, below 18 kHz. Our measurements show that non of the sources interfere with the resonance frequencies used for the transmission.

\begin{table}[]
	\caption{Sources of acoustic interference}
	\label{tab:inter}
	\begin{tabular}{@{}ll@{}}
		\toprule
		Acoustic source     & Max frequency \\ \midrule
		CPU fans             & 525 Hz               \\
		Case fans            & 302 Hz               \\
		PSU fans             & 360 Hz               \\
		Hard-disk-drives (constant, read/write)     & 140 Hz, 8000 Hz             \\
		CD/DVD                & 15000 Hz                 \\
		Humans (speaks, walks)   & 8000 Hz               \\
		Keyboard press        & 8000 Hz (depend on keyboard type)               \\
		Mouse (movements, clicks)      & 14000 Hz - 18000 Hz              \\
		Air-conditioning (all components)    & 4000 Hz               \\ \bottomrule
	\end{tabular}
\end{table}

\section{Countermeasures}
\label{sec:counter}

\textbf{Separation}. Policy-based countermeasures may include the zoning approach, which is used in the telecommunication security standards (e.g., TEMPEST/2-95 \cite{NSTISSAM75:online}). In this approach, systems are kept in restricted zones defined by a different radius, depending on the zone classification. However, this solution is not always feasible due to the practical limitations of available resources. In our case, smartphones should be kept at a range of eight meters or more from the secured area. 

\textbf{Speakers elimination \& blocking}. The transmitting computer exploits the ultrasonic frequencies played through the speakers to generate the resonance. In highly secure facilities, it is common practice to forbid the use of loudspeakers to create an audio-less networking environment known as `audio-gapped' \cite{Jumpingt83:online}. Software countermeasures include removing the audio drivers from the OS or completely disabling the audio hardware in the BIOS level configurations. This can prevent malware from transmitting ultrasonic signals. However, such a configuration eliminates the legitimate use of the audio hardware and is not recommended.

\begin{figure}[h]
	\centering
	\includegraphics[width=0.7\columnwidth]{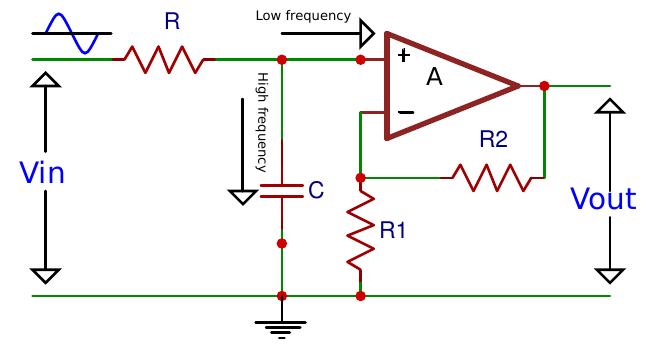}
	\caption{Low-pass filter circuit for the loudspeakers output.}
	\label{fig:lowpass}
\end{figure}

\textbf{Ultrasonic filtering}. It is possible to filter out the resonance frequencies generated by the audio hardware using an audio filter. A software-based ultrasonic firewall like SilverDog and SoniControl can filter the ultrasonic frequencies from websites, media, and streams \cite{GitHubub23:online}. There are open-source projects aimed at blocking device capabilities that are used to generate `ultrasonic beacons' for tracking \cite{mavroudis2017privacy}. A filter could be implemented as an audio stack driver in the operating system to prevent and mask ultrasonic waves. The main drawback of this approach is that it can be disabled or bypassed by sophisticated attacks and kernel-level rootkits. For a hardware level of protection, it was proposed to build a lowpass ultrasonic filter as a trusted hardware component and install it on the loudspeakers \cite{guri2020speaker}. Figure \ref{fig:lowpass} shows the circuit design of an external lowpass filter with an amplifier for a 3.5 mm audio jack. Note that the cutoff frequency in this filter is determined by the capacitor \textit{C} and the resistor \textit{R}.

\begin{figure}[]
	\centering
	\includegraphics[width=1\linewidth]{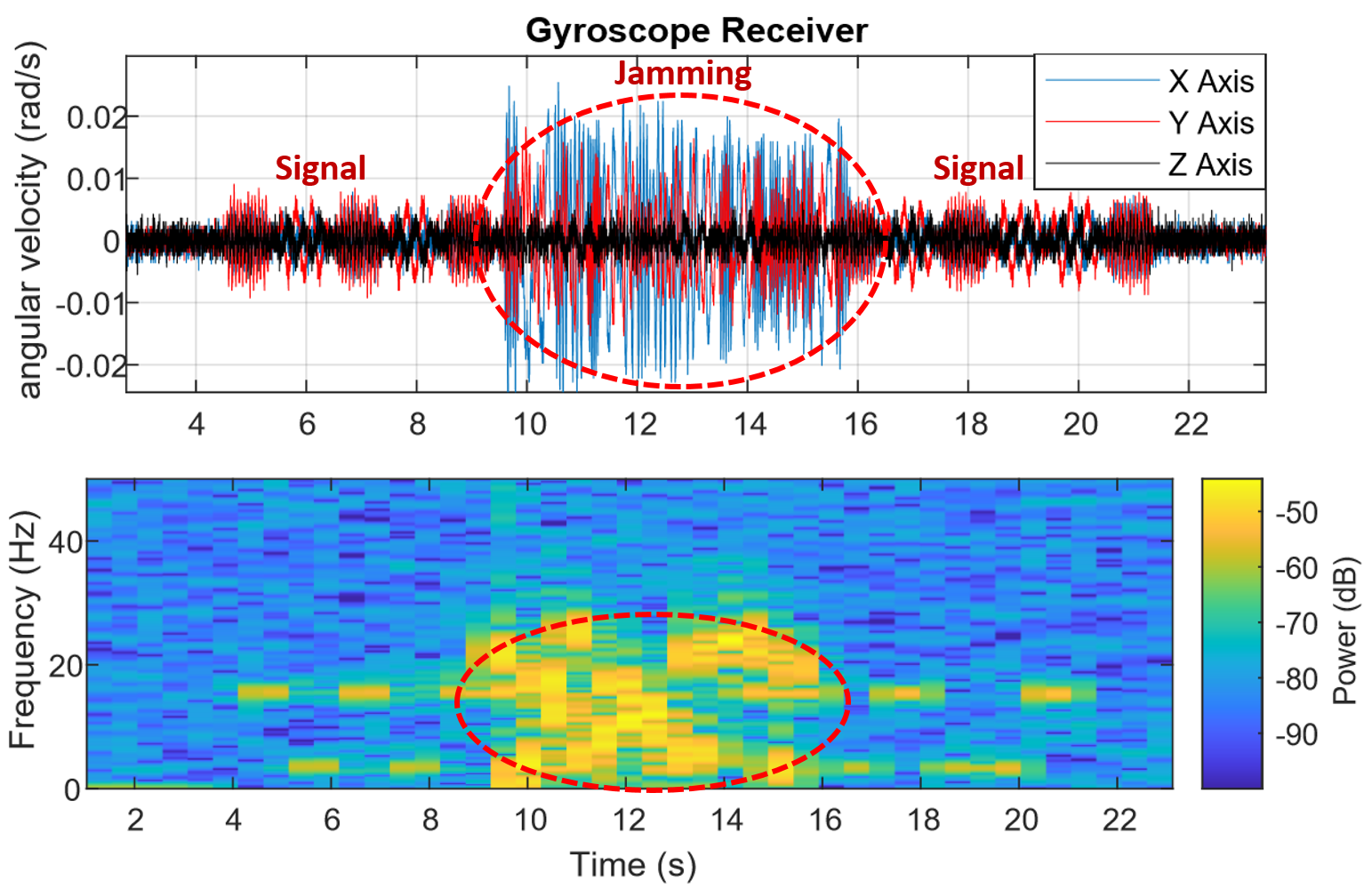}
	\caption{The effect of signal jamming (seconds 10 - 16).}
	\label{fig:jamm1}
\end{figure}

\textbf{Signal jamming}. Another countermeasure method is to jam the covert channel by adding background noises to the acoustic spectrum \cite{AudioJam82:online}. For example, it is possible to mask the original transmissions in a certain area using ultrasonic jammers. These devices generate background noise aimed at interfering with the covert communication signals \cite{9Counter84:online}. In our case, the jammers must be pre-configured with the resonance frequencies of the potential devices. To demonstrate the jamming countermeasures, we built a random ultrasonic white noise generator. The jammer was configured to work at the resonance frequency range and emit random signals at the given frequency band (see Algorithm \ref{alg:a4}). Figure \ref{fig:jamm1} depicts the jammer operation during six seconds. As seen in the waveform and spectrogram, the originally transmitted signal is overridden and disrupted by the random resonance signals from the jammer (seconds 10 - 16). Note that this type of solution is often limited to a single room. Another limitation of this approach is the continuous generation of resonance frequencies that disrupts the gyroscope output.

\begin{algorithm} 
	\caption{ResonanceJammer ($f_{min}$, $f_{max}$, $T$, $maxVolume$)} 
	\begin{algorithmic}[1] 
		\label{alg:4}	
		\State $ rndFreq = random(f_{min}, f_{max}) $
		\State $ rndDuration = random(0, T$)
		\State $ rndVolume = random(0, maxVolume) $
		\State $ vec = GenerateSinWave(rndFreq, T, maxVolume) $
		\State $ playAudio(vec)$ 
		\State $ timeInterval = random(0, T) $
		\State $ sleep(timeInterval)$
		
	\end{algorithmic}
	\label{alg:a4}
\end{algorithm}

\textbf{Signal monitoring}. Some researchers \cite{carrara2014acoustic} suggested monitoring the ultrasonic audio channels for power levels in order to detect convert ultrasound transmissions. Such monitoring products exist \cite{Products93:online} but may produce many false positives due to other interference, and environmental noises in the area \cite{carrara2016survey}.

\textbf{Sensors security}. There are several security frameworks proposed in recent years to minimize the threats of sensor attacks on smartphones.  
Sikder et al. proposed \cite{sikder2019context} 6thSense, a context-aware intrusion detection system that enhances the security of smart devices by observing changes in sensor data.  Xu et al. presented SemaDroid \cite{xu2015semadroid}, a framework that provides comprehensive and fine-grained access control over various sensors. Petracca et al. introduced AuDroid \cite{petracca2015audroid}, an extension to the SELinux reference that monitors and enforces the use of system audio resources such as microphones. 
Note that all these protections can be bypassed by using device vulnerabilities and OS exploits. For example, major vulnerabilities disclosed in 2021 (e.g., CVE-2021-1905 and CVE-2021-1906) enable the execution of malicious code on Android with root privileges and give them full control of the device  \cite{4vulnera80:online}.

\textbf{Attack prevention}. In order to eliminate the interferences from external acoustic sources is was proposed to shield the gyroscope circuit from acoustic noise. This is effective would prevent acoustic attacks. However, covering the sensor with a sound blocking case is not always feasible, such as a small MEMS gyroscope integrated into smartphones. In systems sensitive to spoofing and injection attacks (e.g., UAVs), it is possible to use non-mechanical sensors, such as gas-bearing gyroscopes and optical gyroscopes, which are not susceptible to acoustic attacks.

\section{Conclusion}
\label{sec:conclusion}
This paper shows that malware running on an air-gapped computer can transmit data to a nearby smartphone via ultrasonic frequencies. However, to receive the ultrasonic signals, we are not using the microphone, a highly protected sensor in the mobile OS (e.g., Android and iOS). Instead, we produce inaudible tones in specific resonance frequencies that generate tiny vibrations within the gyroscope sensor. Data is modulated on these resonance frequencies and then decoded via the vibrations generated in the nearby smartphone. Notably, the gyroscope in mobile devices requires only low permissions because it is considered a safe sensor.
We presented the attack model and provided related technical background. We introduced the transmitter and receiver and discussed their design and implementation. We evaluated the covert channel and showed that data could be exfiltrated from air-gapped computers to nearby smartphones to distances of several meters away via  Speakers-to-Gyroscope communication.


\balance
\bibliographystyle{plain}
\bibliography{../../AirGap,../../AirGapCases,../../mobile,../../AirGapTools,../../PowerSupply,../../vibrations,gyro}

\begin{thebibliography}{10}

\bibitem{4vulnera80:online}
4 vulnerabilities under attack give hackers full control of android devices |
  ars technica.
\newblock
  \url{https://arstechnica.com/gadgets/2021/05/hackers-have-been-exploiting-4-critical-android-vulnerabilities/}.
\newblock (Accessed on 09/05/2022).

\bibitem{AgentBTZ65:online}
Agent.btz - wikipedia.
\newblock \url{https://en.wikipedia.org/wiki/Agent.BTZ}.
\newblock (Accessed on 09/05/2022).

\bibitem{ESETRese49:online}
Eset research discovers cyber espionage framework ramsay | eset.
\newblock
  \url{https://www.eset.com/int/about/newsroom/press-releases/research/eset-research-discovers-cyber-espionage-framework-ramsay/}.
\newblock (Accessed on 09/05/2022).

\bibitem{Hackerst65:online}
Hackers target the air-gapped networks of the taiwanese and philippine military
  | zdnet.
\newblock
  \url{https://www.zdnet.com/article/hackers-target-the-air-gapped-networks-of-the-taiwanese-and-philippine-military/}.
\newblock (Accessed on 09/05/2022).

\bibitem{Microsof69:online}
Microsoft launches 'top secret' azure cloud region for us intelligence
  community.
\newblock
  \url{https://www.computerweekly.com/news/252505336/Microsoft-launches-top-secret-Azure-cloud-region-for-US-intelligence-community}.
\newblock (Accessed on 09/05/2022).

\bibitem{Nobigdea65:online}
No big deal... kremlin hackers 'jumped air-gapped networks' to pwn us power
  utilities - the register.
\newblock
  \url{https://www.theregister.co.uk/2018/07/24/russia_us_energy_grid_hackers/}.
\newblock (Accessed on 09/05/2022).

\bibitem{Sensoron5:online}
Sensor.onreading - web apis | mdn.
\newblock
  \url{https://developer.mozilla.org/en-US/docs/Web/API/Sensor/onreading}.
\newblock (Accessed on 11/05/2021).

\bibitem{Storefro90:online}
Storefront - top secret/sensitive compartmented information data.
\newblock
  \url{https://storefront.disa.mil/kinetic/disa/service-catalog#/forms/top-secretsensitive-compartmented-information-data}.
\newblock (Accessed on 09/05/2022).

\bibitem{9Counter84:online}
9 counter surveillance tools you can legally use | independent living news.
\newblock
  \url{https://independentlivingnews.com/2013/11/12/20397-9-counter-surveillance-tools-you-can-legally-use/},
  2013.
\newblock (Accessed on 09/05/2022).

\bibitem{funtenna86:online}
funtenna - github.
\newblock \url{https://github.com/funtenna}, 2016.
\newblock (Accessed on 09/05/2022).

\bibitem{GitHubub23:online}
Github - ubeacsec/silverdog: An audio firewall for chrome!
\newblock \url{https://github.com/ubeacsec/Silverdog}, 2017.
\newblock (Accessed on 09/05/2022).

\bibitem{AudioJam82:online}
Audio jammer | counter surveillance systems.
\newblock
  \url{https://www.brickhousesecurity.com/counter-surveillance/audio-jammers/},
  2018.
\newblock (Accessed on 09/05/2022).

\bibitem{Praatdoi67:online}
Praat: doing phonetics by computer.
\newblock \url{http://www.fon.hum.uva.nl/praat/}, 2018.
\newblock (Accessed on 09/05/2022).

\bibitem{Products93:online}
Products - pulsar instruments plc.
\newblock \url{https://pulsarinstruments.com/en/categories}, 2018.
\newblock (Accessed on 09/05/2022).

\bibitem{algamili2021review}
Abdullah~Saleh Algamili, Mohd Haris~Md Khir, John~Ojur Dennis, Abdelaziz~Yousif
  Ahmed, Sami~Sultan Alabsi, Saeed Salem~Ba Hashwan, and Mohammed~M Junaid.
\newblock A review of actuation and sensing mechanisms in mems-based sensor
  devices.
\newblock {\em Nanoscale research letters}, 16(1):1--21, 2021.

\bibitem{byres2013air}
Eric Byres.
\newblock The air gap: Scada's enduring security myth.
\newblock {\em Communications of the ACM}, 56(8):29--31, 2013.

\bibitem{carrara2014acoustic}
Brent Carrara and Carlisle Adams.
\newblock On acoustic covert channels between air-gapped systems.
\newblock In {\em International Symposium on Foundations and Practice of
  Security}, pages 3--16. Springer, 2014.

\bibitem{carrara2016survey}
Brent Carrara and Carlisle Adams.
\newblock A survey and taxonomy aimed at the detection and measurement of
  covert channels.
\newblock In {\em Proceedings of the 4th ACM Workshop on Information Hiding and
  Multimedia Security}, pages 115--126. ACM, 2016.

\bibitem{cova2010detection}
Marco Cova, Christopher Kruegel, and Giovanni Vigna.
\newblock Detection and analysis of drive-by-download attacks and malicious
  javascript code.
\newblock In {\em Proceedings of the 19th international conference on World
  wide web}, pages 281--290. ACM, 2010.

\bibitem{AnIndian12:online}
Debak Das.
\newblock An indian nuclear power plant suffered a cyberattack. here’s what
  you need to know. - the washington post (04/11/2019).
\newblock \url{https://www.washingtonpost.com}.

\bibitem{Jumpingt83:online}
Josh Dean.
\newblock Jumping the airgap.
\newblock \url{https://thoughtworksnc.com/2017/03/16/jumping-the-airgap/}, 03
  2017.
\newblock (Accessed on 09/05/2022).

\bibitem{deshotels2014inaudible}
Luke Deshotels.
\newblock Inaudible sound as a covert channel in mobile devices.
\newblock In {\em WOOT}, 2014.

\bibitem{Deshotels2014}
Luke Deshotels.
\newblock Inaudible sound as a covert channel in mobile devices.
\newblock In {\em WOOT}, 2014.

\bibitem{giani2006data}
Annarita Giani, Vincent~H Berk, and George~V Cybenko.
\newblock Data exfiltration and covert channels.
\newblock In {\em Defense and Security Symposium}, pages 620103--620103.
  International Society for Optics and Photonics, 2006.

\bibitem{guri2020cd}
Mordechai Guri.
\newblock Cd-leak: Leaking secrets from audioless air-gapped computers using
  covert acoustic signals from cd/dvd drives.
\newblock In {\em 2020 IEEE 44th Annual Computers, Software, and Applications
  Conference (COMPSAC)}, pages 808--816. IEEE, 2020.

\bibitem{guri2020power}
Mordechai Guri.
\newblock Power-supplay: Leaking data from air-gapped systems by turning the
  power-supplies into speakers.
\newblock {\em arXiv preprint arXiv:2005.00395}, 2020.

\bibitem{guri2019air}
Mordechai Guri and Dima Bykhovsky.
\newblock air-jumper: Covert air-gap exfiltration/infiltration via security
  cameras \& infrared (ir).
\newblock {\em Computers \& Security}, 82:15--29, 2019.

\bibitem{guri2019brightness}
Mordechai Guri, Dima Bykhovsky, and Yuval Elovici.
\newblock Brightness: Leaking sensitive data from air-gapped workstations via
  screen brightness.
\newblock In {\em 2019 12th CMI Conference on Cybersecurity and Privacy (CMI)},
  pages 1--6. IEEE, 2019.

\bibitem{guri2018magneto}
Mordechai Guri, Andrey Daidakulov, and Yuval Elovici.
\newblock Magneto: Covert channel between air-gapped systems and nearby
  smartphones via cpu-generated magnetic fields.
\newblock {\em arXiv preprint arXiv:1802.02317}, 2018.

\bibitem{Guri:2018:BAM:3200906.3177230}
Mordechai Guri and Yuval Elovici.
\newblock Bridgeware: The air-gap malware.
\newblock {\em Commun. ACM}, 61(4):74--82, March 2018.

\bibitem{guri2016optical}
Mordechai Guri, Ofer Hasson, Gabi Kedma, and Yuval Elovici.
\newblock An optical covert-channel to leak data through an air-gap.
\newblock In {\em Privacy, Security and Trust (PST), 2016 14th Annual
  Conference on}, pages 642--649. IEEE, 2016.

\bibitem{guri2015gsmem}
Mordechai Guri, Assaf Kachlon, Ofer Hasson, Gabi Kedma, Yisroel Mirsky, and
  Yuval Elovici.
\newblock Gsmem: Data exfiltration from air-gapped computers over gsm
  frequencies.
\newblock In {\em USENIX Security Symposium}, pages 849--864, 2015.

\bibitem{guri2014airhopper}
Mordechai Guri, Gabi Kedma, Assaf Kachlon, and Yuval Elovici.
\newblock Airhopper: Bridging the air-gap between isolated networks and mobile
  phones using radio frequencies.
\newblock In {\em Malicious and Unwanted Software: The Americas (MALWARE), 2014
  9th International Conference on}, pages 58--67. IEEE, 2014.

\bibitem{Guri2016}
Mordechai Guri, Matan Monitz, and Yuval Elovici.
\newblock Usbee: Air-gap covert-channel via electromagnetic emission from usb.
\newblock In {\em Privacy, Security and Trust (PST), 2016 14th Annual
  Conference on}, pages 264--268. IEEE, 2016.

\bibitem{guri2016usbee}
Mordechai Guri, Matan Monitz, and Yuval Elovici.
\newblock Usbee: Air-gap covert-channel via electromagnetic emission from usb.
\newblock In {\em Privacy, Security and Trust (PST), 2016 14th Annual
  Conference on}, pages 264--268. IEEE, 2016.

\bibitem{guri2017bridging}
Mordechai Guri, Matan Monitz, and Yuval Elovici.
\newblock Bridging the air gap between isolated networks and mobile phones in a
  practical cyber-attack.
\newblock {\em ACM Transactions on Intelligent Systems and Technology (TIST)},
  8(4):50, 2017.

\bibitem{guri2015bitwhisper}
Mordechai Guri, Matan Monitz, Yisroel Mirski, and Yuval Elovici.
\newblock Bitwhisper: Covert signaling channel between air-gapped computers
  using thermal manipulations.
\newblock In {\em Computer Security Foundations Symposium (CSF), 2015 IEEE
  28th}, pages 276--289. IEEE, 2015.

\bibitem{guri2017acoustic}
Mordechai Guri, Yosef Solewicz, Andrey Daidakulov, and Yuval Elovici.
\newblock Acoustic data exfiltration from speakerless air-gapped computers via
  covert hard-drive noise (diskfiltration).
\newblock In {\em European Symposium on Research in Computer Security}, pages
  98--115. Springer, 2017.

\bibitem{Guri2018Mosquito}
Mordechai Guri, Yosef Solewicz, and Yuval Elovici.
\newblock Mosquito: Covert ultrasonic transmissions between two air-gapped
  computers using speaker-to-speaker communication.
\newblock In {\em 2018 IEEE Conference on Dependable and Secure Computing
  (DSC)}, pages 1--8. IEEE, 2018.

\bibitem{guri2020fansmitter}
Mordechai Guri, Yosef Solewicz, and Yuval Elovici.
\newblock Fansmitter: Acoustic data exfiltration from air-gapped computers via
  fans noise.
\newblock {\em Computers \& Security}, page 101721, 2020.

\bibitem{guri2020speaker}
Mordechai Guri, Yosef Solewicz, and Yuval Elovici.
\newblock Speaker-to-speaker covert ultrasonic communication.
\newblock {\em Journal of Information Security and Applications}, 51:102458,
  2020.

\bibitem{guri2019ctrl}
Mordechai Guri, Boris Zadov, Dima Bykhovsky, and Yuval Elovici.
\newblock Ctrl-alt-led: Leaking data from air-gapped computers via keyboard
  leds.
\newblock In {\em 2019 IEEE 43rd Annual Computer Software and Applications
  Conference (COMPSAC)}, volume~1, pages 801--810. IEEE, 2019.

\bibitem{guri2019powerhammer}
Mordechai Guri, Boris Zadov, Dima Bykhovsky, and Yuval Elovici.
\newblock Powerhammer: Exfiltrating data from air-gapped computers through
  power lines.
\newblock {\em IEEE Transactions on Information Forensics and Security}, 2019.

\bibitem{guri2018xled}
Mordechai Guri, Boris Zadov, Andrey Daidakulov, and Yuval Elovici.
\newblock xled: Covert data exfiltration from air-gapped networks via switch
  and router leds.
\newblock In {\em 2018 16th Annual Conference on Privacy, Security and Trust
  (PST)}, pages 1--12. IEEE, 2018.

\bibitem{Guri2017}
Mordechai Guri, Boris Zadov, and Yuval Elovici.
\newblock {\em LED-it-GO: Leaking (A Lot of) Data from Air-Gapped Computers via
  the (Small) Hard Drive LED}, pages 161--184.
\newblock Springer International Publishing, Cham, 2017.

\bibitem{guri2019odini}
Mordechai Guri, Boris Zadov, and Yuval Elovici.
\newblock Odini: Escaping sensitive data from faraday-caged, air-gapped
  computers via magnetic fields.
\newblock {\em IEEE Transactions on Information Forensics and Security},
  15:1190--1203, 2019.

\bibitem{hanspach2014covert}
Michael Hanspach and Michael Goetz.
\newblock On covert acoustical mesh networks in air.
\newblock {\em arXiv preprint arXiv:1406.1213}, 2014.

\bibitem{hasan2013sensing}
Ragib Hasan, Nitesh Saxena, Tzipora Haleviz, Shams Zawoad, and Dustin Rinehart.
\newblock Sensing-enabled channels for hard-to-detect command and control of
  mobile devices.
\newblock In {\em Proceedings of the 8th ACM SIGSAC symposium on Information,
  computer and communications security}, pages 469--480, 2013.

\bibitem{NSTISSAM75:online}
https://cryptome.org.
\newblock Nstissam tempest/2-95.
\newblock \url{https://cryptome.org/tempest-2-95.htm}, 2000.
\newblock (Accessed on 09/05/2022).

\bibitem{khazaaleh2019vulnerability}
Shadi Khazaaleh, Georgios Korres, Mohammed Eid, Mahmoud Rasras, and Mohammed~F
  Daqaq.
\newblock Vulnerability of mems gyroscopes to targeted acoustic attacks.
\newblock {\em IEEE Access}, 7:89534--89543, 2019.

\bibitem{kuhn1998soft}
Markus~G Kuhn and Ross~J Anderson.
\newblock Soft tempest: Hidden data transmission using electromagnetic
  emanations.
\newblock In {\em Information hiding}, volume 1525, pages 124--142. Springer,
  1998.

\bibitem{kuhn2002compromising}
Markus~Guenther Kuhn.
\newblock {\em Compromising emanations: eavesdropping risks of computer
  displays}.
\newblock PhD thesis, University of Cambridge, 2002.

\bibitem{kushner2013real}
David Kushner.
\newblock The real story of stuxnet.
\newblock {\em ieee Spectrum}, 3(50):48--53, 2013.

\bibitem{loughry2002information}
Joe Loughry and David~A Umphress.
\newblock Information leakage from optical emanations.
\newblock {\em ACM Transactions on Information and System Security (TISSEC)},
  5(3):262--289, 2002.

\bibitem{madhavapeddy2005audio}
Anil Madhavapeddy, Richard Sharp, David Scott, and Alastair Tse.
\newblock Audio networking: the forgotten wireless technology.
\newblock {\em IEEE Pervasive Computing}, 4(3):55--60, 2005.

\bibitem{matyunin2019vibrational}
Nikolay Matyunin, Yujue Wang, and Stefan Katzenbeisser.
\newblock Vibrational covert channels using low-frequency acoustic signals.
\newblock In {\em Proceedings of the ACM Workshop on Information Hiding and
  Multimedia Security}, pages 31--36, 2019.

\bibitem{mavroudis2017privacy}
Vasilios Mavroudis, Shuang Hao, Yanick Fratantonio, Federico Maggi, Christopher
  Kruegel, and Giovanni Vigna.
\newblock On the privacy and security of the ultrasound ecosystem.
\newblock {\em Proceedings on Privacy Enhancing Technologies}, 2017(2):95--112,
  2017.

\bibitem{murdoch2005embedding}
Steven~J Murdoch and Stephen Lewis.
\newblock Embedding covert channels into tcp/ip.
\newblock In {\em Information hiding}, volume 3727, pages 247--261. Springer,
  2005.

\bibitem{peltier2006social}
Thomas~R Peltier.
\newblock Social engineering: Concepts and solutions.
\newblock {\em Information Systems Security}, 15(5):13--21, 2006.

\bibitem{petracca2015audroid}
Giuseppe Petracca, Yuqiong Sun, Trent Jaeger, and Ahmad Atamli.
\newblock Audroid: Preventing attacks on audio channels in mobile devices.
\newblock In {\em Proceedings of the 31st Annual Computer Security Applications
  Conference}, pages 181--190, 2015.

\bibitem{provos2007ghost}
Niels Provos, Dean McNamee, Panayiotis Mavrommatis, Ke~Wang, Nagendra Modadugu,
  et~al.
\newblock The ghost in the browser: Analysis of web-based malware.
\newblock {\em HotBots}, 7:4--4, 2007.

\bibitem{sikder2019context}
Amit~Kumar Sikder, Hidayet Aksu, and A~Selcuk Uluagac.
\newblock A context-aware framework for detecting sensor-based threats on smart
  devices.
\newblock {\em IEEE Transactions on Mobile Computing}, 19(2):245--261, 2019.

\bibitem{smutz2012malicious}
Charles Smutz and Angelos Stavrou.
\newblock Malicious pdf detection using metadata and structural features.
\newblock In {\em Proceedings of the 28th annual computer security applications
  conference}, pages 239--248. ACM, 2012.

\bibitem{son2015rocking}
Yunmok Son, Hocheol Shin, Dongkwan Kim, Youngseok Park, Juhwan Noh, Kibum Choi,
  Jungwoo Choi, and Yongdae Kim.
\newblock Rocking drones with intentional sound noise on gyroscopic sensors.
\newblock In {\em 24th $\{$USENIX$\}$ Security Symposium ($\{$USENIX$\}$
  Security 15)}, pages 881--896, 2015.

\bibitem{sood2011malvertising}
Aditya~K Sood and Richard~J Enbody.
\newblock Malvertising--exploiting web advertising.
\newblock {\em Computer Fraud \& Security}, 2011(4):11--16, 2011.

\bibitem{tu2018injected}
Yazhou Tu, Zhiqiang Lin, Insup Lee, and Xiali Hei.
\newblock Injected and delivered: Fabricating implicit control over actuation
  systems by spoofing inertial sensors.
\newblock In {\em 27th $\{$USENIX$\}$ Security Symposium ($\{$USENIX$\}$
  Security 18)}, pages 1545--1562, 2018.

\bibitem{vuagnoux2009compromising}
Martin Vuagnoux and Sylvain Pasini.
\newblock Compromising electromagnetic emanations of wired and wireless
  keyboards.
\newblock In {\em USENIX security symposium}, pages 1--16, 2009.

\bibitem{xu2015semadroid}
Zhi Xu and Sencun Zhu.
\newblock Semadroid: A privacy-aware sensor management framework for
  smartphones.
\newblock In {\em Proceedings of the 5th ACM Conference on Data and Application
  Security and Privacy}, pages 61--72, 2015.

\bibitem{yonezawa2011transferring}
Takuro Yonezawa, Tomotaka Ito, and Hideyuki Tokuda.
\newblock Transferring information from mobile devices to personal computers by
  using vibration and accelerometer.
\newblock In {\em Proceedings of the 13th international conference on
  Ubiquitous computing}, pages 487--488, 2011.

\bibitem{zander2007survey}
Sebastian Zander, Grenville Armitage, and Philip Branch.
\newblock A survey of covert channels and countermeasures in computer network
  protocols.
\newblock {\em IEEE Communications Surveys \& Tutorials}, 9(3):44--57, 2007.

\end{thebibliography}


\end{document}